\documentclass[%
prd,
reprint,nodate,
superscriptaddress,
nofootinbib,
 amsmath,amssymb,
 aps,
]{revtex4-1}

\pdfoutput=1

\usepackage[utf8]{inputenc} 

\usepackage{xcolor}
 
\usepackage{graphicx}
\usepackage{dcolumn}
\usepackage{bm}
\usepackage{microtype}
\usepackage{threeparttable} 
\usepackage{mathtools}
\usepackage[colorlinks=true]{hyperref}

\newcommand{\be}{\begin{eqnarray}}
\newcommand{\ee}{\end{eqnarray}}

\begin{document}
\title{Chaotic internal dynamics of dissipative optical soliton molecules}

    \author{Youjian Song}
\thanks{These authors contributed equally to this work.}
\author{Defeng Zou}
\thanks{These authors contributed equally to this work.}
\affiliation{%
	Ultrafast Laser Laboratory, Key Laboratory of Opto-electronic Information Science and Technology of Ministry of Education, School of Precision Instruments and Opto-electronics Engineering, Tianjin University, Tianjin 300072, China}%

\author{Omri Gat}%
\email{omrigat@mail.huji.ac.il}
\affiliation{%
	Racah Institute of Physics, Hebrew University of Jerusalem, Jerusalem 91904, Israel}

\author{Minglie Hu}%
\email{huminglie@tju.edu.cn}
\affiliation{%
	Ultrafast Laser Laboratory, Key Laboratory of Opto-electronic Information Science and Technology of Ministry of Education, School of Precision Instruments and Opto-electronics Engineering, Tianjin University, Tianjin 300072, China}

\author{Philippe Grelu}%
\email{philippe.grelu@u-bourgogne.fr}
\affiliation{%
	Laboratoire ICB UMR 6303 CNRS, Université Bourgogne Franche-Comté, 9 avenue A. Savary, Dijon 21000, France}

\begin{abstract}
When a laser cavity supports the propagation of several ultrashort pulses, these pulses interact and can form compact bound states called soliton molecules. Soliton molecules are fascinating objects of nonlinear science, which present striking analogies with their matter molecules counterparts. The soliton pair, composed of two identical pulses, constitutes the chief soliton molecule of fundamental interest. The relative timing and phase between the two propagating pulses are the most salient internal degrees of freedom of the soliton molecule. These two internal degrees of freedom allow self-oscillating soliton molecules, which have indeed been repeatedly observed, whereas the low-dimensional chaotic dynamics of a soliton-pair molecule remains elusive, noting that it would require at least three degrees of freedom. We here report the observation of chaotic soliton-pair molecules within an ultrafast fiber laser, by means of a direct measurement of the relative optical pulse separation with sub-femtosecond precision in real time. Moreover, we demonstrate an all-optical control of the chaotic dynamics followed by the soliton molecule, by injecting a modulated optical signal that resynchronizes the internal periodic vibration of soliton molecule. 
\end{abstract}

\maketitle

\section{Introduction}\label{sec:introduction}

Chaotic dynamics are characterized by aperiodicity, high sensitivity to initial conditions and long-term unpredictability. These are intrinsic features and behaviors observable in most nonlinear systems. The manifestation of chaos in extended natural systems raises awe and fear, from the uncertain evolution of celestial bodies to the weather instability, respectively highlighted in the seminal works of Poincaré and Lorenz \cite{poincare1890probleme,lorenz1963deterministic}. Therefore, from the past century, there has been a high impetus to investigate chaos within laboratory-confined systems over a wide range of scientific fields, from hydrodynamics and electromagnetics to quantum physics and optics \cite{strogatz2018nonlinear}. Studying chaos in nonlinear optics is of particular interest since photonics exhibits handy systems such as the laser oscillator that can advance the fundamental understanding of chaotic dynamics and utilize chaos in technological innovations such as in secure optical communications and Lidar sensing in automotive environments \cite{argyris2005chaos,uchida2008fast,spitz2021private}. Alternatively, understanding the early warnings signaling the onset of chaos helps circumvent its adverse manifestations in laser applications requiring high stability \cite{sciamanna2015physics,jiang2017chaos,jumpertz2016chaotic,fan2021real,deng2022mid,steinmeyer2021chaotic}.

The ultrafast laser constitutes an outstanding system for the investigation of chaos, being inherently wideband, highly nonlinear, and offering multiple parameter control schemes. Furthermore, the commonly accepted “dissipative soliton” paradigm explicates the great diversity of ultrashort laser pulse dynamics, which spans from stationary to chaotic pulse emission. Dissipative solitons are attracting states that result from the composite balance between dissipative and dispersive propagation effects and are potentially subjected to a great variety of bifurcations \cite{grelu2012dissipative}. The standard mode-locked laser regime features a single laser pulse that retakes the same evolution along successive cavity roundtrips: this results from the existence of a stable focus attractor providing long term stability and robustness. Within the parameter space of the optical cavity, the laser can undergo abrupt bifurcations leading to chaotic pulse dynamics \cite{sucha1995period,xing1999regular,soto2004bifurcations,zhao2006chaotic}. Vivid illustrations of these chaotic dynamics include “soliton explosions” \cite{cundiff2002experimental,runge2015observation}, which manifest as intermittent chaos in the pulse evolution over numerous cavity roundtrips, and “noise-like” pulse dynamics, characterized by a persisting noisy optical waveform and a complete loss of mutual coherence between successive roundtrips \cite{horowitz1997noiselike,runge2013coherence,lecaplain2014rogue}. Soliton explosions and noise-like pulsing belong to the category of “incoherent dissipative solitons”, which involves chaotic attractors to solve the apparent paradox of an average temporal localization for the pulse and its high instability \cite{krupa2017vector,soto2005soliton,wang2020buildup}. We emphasize that these chaotic solitons evolve around attractors of extended dynamical systems featuring a high dimensionality, with similar possibilities within Kerr microresonators \cite{lucas2017breathing}, noting that the existence of low-dimensional chaotic soliton dynamics has been recently reported \cite{melo2018deterministic,wang2022real}.

Other abrupt bifurcations may result from the laser multi-pulsing instability (MPI) \cite{shlizerman2011characterizing}. Typically, when the pump power is augmented, the MPI leads to an increased number of intracavity pulses. Therefore, crossing the MPI with subsequent complex bifurcations leads to a virtually unlimited landscape of complex and chaotic ultrafast multi-pulse dynamics \cite{lecaplain2012dissipative,chouli2010soliton,zaviyalov2012rogue,liu2016successive,meng2021intracavity}. The interaction between dissipative solitons is also known to lead to the spontaneous formation of compact bound states, also known as dissipative soliton molecules, since they display striking analogies with their matter molecule counterparts \cite{grelu2012dissipative}. Akin to the hydrogen molecule, the soliton-pair molecule is made of two identical constituents: it is the chief soliton molecule of fundamental interest, and the focus of our present investigations. Once formed, a stable optical soliton molecule can propagate indefinitely within the laser cavity. For a soliton molecule being stationary in its moving frame, the relative temporal separation and phase between the soliton constituents remain constant. Stationary soliton molecules are routinely generated in laser oscillators \cite{soto2003quantized,tang2005mechanism,wang2017universal,song2020attosecond} and can be represented as point attractors in an infinite-dimensional phase space. However, upon a change in the laser parameters, a Hopf-type bifurcation can take place and generate pulsating soliton molecules, represented as limit-cycle attractors, see Fig.1 (c) \cite{grapinet2006vibrating,soto2007soliton}. These self-excited oscillations with time-varying properties can manifest in various ways, encompassing vibrating soliton molecules – where the soliton separation oscillates – as well as phase-only oscillations \cite{ortacc2010observation,krupa2017real,herink2017real,liu2018real,wang2019optical,melchert2019soliton,luo2020real,zhou2020buildup,peng2021breather,du2022internal,nimmesgern2021soliton}. Among periodic oscillations, there exist various relative soliton trajectories, from quasi-harmonic to strongly anharmonic \cite{hamdi2018real,hamdi2022superlocalization}. 

Recent laser experiments have implemented strategies of external periodic perturbation to probe and even induce transitions within soliton molecules, bringing in a close analogy to the approach of molecular spectroscopy. For instance, the resonant excitation of optical soliton molecules within a Kerr-lens mode-locked Ti:sapphire laser probed the intramolecular interaction to reveal its anharmonic feature \cite{kurtz2020resonant}. The internal oscillatory dynamics of optical soliton molecules have been experimentally synchronized with a modulated signal injected into an ultrafast fiber laser \cite{zou2022synchronization}. These experiments strengthen the hypothesis that the individual solitons composing a multisoliton waveform are often governed by low-dimensional dynamical systems. Since optical solitons break phase symmetry as well as time-translation symmetry, the internal dynamics of a soliton-pair molecule has at least two degrees of freedom, which is enough for self-oscillations but not for chaos. As a matter of fact, there has not been any observation to date of low-dimensional chaos concerning the internal motions within soliton molecules, either in theory or in experiment. The seminal theoretical predictions of chaotic soliton-pair molecules, based on the complex Ginzburg-Landau equation, needed either the addition of an external drive \cite{turaev2007chaotic} or a soliton shape instability \cite{soto2007soliton}. 

The motivation for discovering low-dimensional chaotic soliton molecules is two-fold. It will strengthen the analogy with the molecules of matter, which are indeed subjected to nonlinear and chaotic vibrations \cite{wu2005nonlinearity}. In addition, it will stimulate the development of effective low-dimensional interaction models applicable within a given range of system parameters inside otherwise highly multi-dimensional complex systems, improving the prospects of analysis and control of optical soliton molecules. 

Hence, in this article, we demonstrate the excitation and all-optical control of a chaotic soliton-pair molecule within a passively mode locked fiber laser. Such experiment requires an extremely well-resolved recording of the relative soliton separation in real-time, which is achieved with the balanced optical cross-correlation (BOC) technique \cite{kim2007attosecond,jia2020photonic}. Indeed, BOC tracks the pulse separation variations within a vibrating soliton molecule with sub-femtosecond resolution, enabling the observation of its transitions to chaotic dynamics for the first time, to the best of our knowledge. The route through period doubling bifurcations to chaos is explicitly retrieved with a perfect reproducibility. The chaotic dynamics are qualitatively characterized by the direct BOC waveform, its radiofrequency (RF) spectrogram and phase portraits analysis. They are also quantitatively analyzed using the Lyapunov exponent and the correlation dimension analysis. Finally, we achieve an all-optical control of the chaotic dynamics followed by the soliton molecule, by means of a weak external signal injection, without altering the central laser parameters. The experimental observations are qualitatively well reproduced by numerical simulations.

\section{Experimental results and analysis}
\subsection{Monitoring the temporal extension of the soliton molecule in real time}

Figure 1 displays the concept and implementation of the characterization of dissipative optical soliton molecules. These molecules are generated within a fiber ring laser cavity, which comprises an erbium-doped fiber for laser emission around the 1.5-micron telecom wavelength, and an effective and controllable ultrafast saturable absorber based on the nonlinear polarization evolution that takes place in the single-mode optical fibers (see the Appendix Section). Upon the adjustment of the pump power and the intracavity retarding waveplates, the laser can self-generate soliton-pair molecules in various dynamical states. The output soliton-pair molecules are beam-split into two branches and recombined with orthogonal polarization states using a Michelson interferometer, as sketched in Fig. 1a. We then detect the real time dynamics of the variations of the pulse separation within the soliton molecule by means of the BOC measurement technique. When the overlap between the two arms is scanned, the BOC outputs an S-shaped voltage signal (see Fig. 1b). We set the zero crossing of the S-shaped curve as the working point, around which the voltage excursions of the BOC output are proportional to the variations of the pulse separation within the soliton molecule, whereas the laser intensity noise is cancelled out by the balanced photodetection (more details about the BOC measurement technique are provided in Supplementary Section A).

\begin{figure*}
	\begin{center}
    \includegraphics[width=11.2cm]{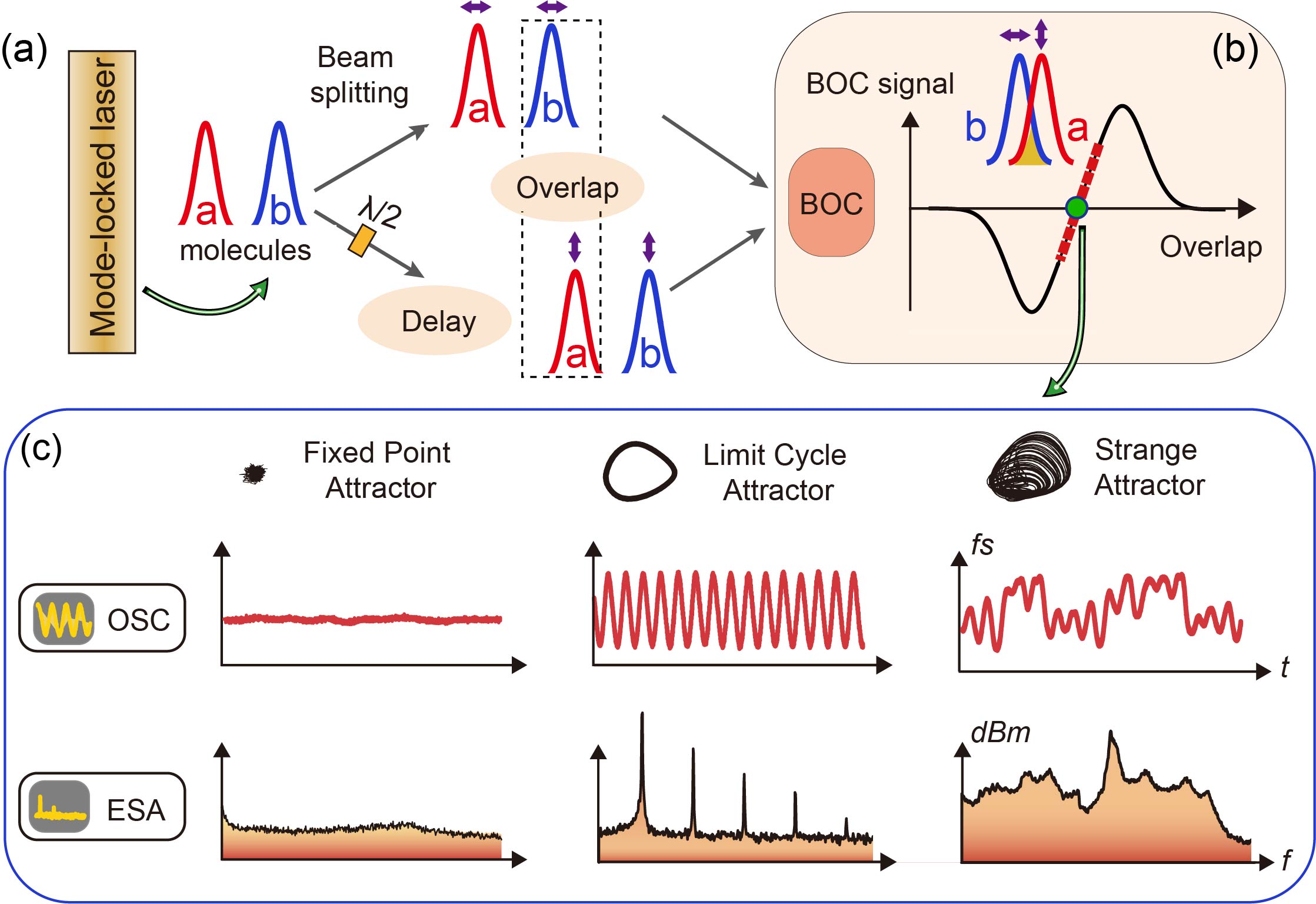}
	\caption{Schematic diagram of the real-time characterization of the extension of dissipative optical soliton molecules. (a) The soliton molecules with diverse internal dynamics are generated by a mode-locked fiber laser and sent into the BOC measurement. (b) An S-shaped curve is obtained when the delay between the two arms is scanned, which allows to define a linear measurement range (red-dotted line). (c) We could measure three intra-molecular dynamics: stationary states with a fixed separation, vibrating states following a regular oscillation, and chaotic states characterized by unpredictable separation dynamics. They respectively correspond to: fixed points, limit cycles and strange attractors of the effective dynamical system. OSC: oscilloscope. ESA: electric spectrum analyzer.}
    \end{center}
\end{figure*}

The recording of the BOC voltage signal allows true real time monitoring of the pulse separation at the sub-femtosecond level without relying on any heavy data post-processing. Supported with a spectral analysis, such monitoring facilitates a comprehensive exploration of complex and low-amplitude internal soliton molecule dynamics. In the case of stationary soliton molecules, the BOC output displays a flat curve with small random fluctuations, whereas for vibrating molecules, periodic trajectories are observed. We anticipate that a breakdown of the periodicity of the time-domain BOC signal can be, in case significant excursions of the BOC signal remain, a signature of chaotic dynamics within the soliton molecule (see Fig. 1c).

\subsection{Generating chaotic soliton molecules}
We have developed a controllable way to excite soliton molecules through a cascade of period-doubling bifurcations leading to chaos \cite{ott2002chaos}. Figure 2a is an extensive recording of the evolution of the RF spectrogram of the BOC output when the pump strength is gradually increased. As a starting point, we set a fundamental period-one (P1) oscillating soliton molecule. An example is shown in the lower part of Fig. 2b, featuring quasi-sinusoidal oscillations of the pulse separation. At a pump driving current $I$ $=$ 509 mA (optical pump power 305 mW), the oscillation of the BOC output signal corresponds to an intramolecular vibration between the two solitons having an amplitude of $\sim$ 80 fs. The oscillation frequency $f_{s}$ is measured from the corresponding RF spectrum displayed in Fig. 2c, featuring a sharp peak located at 2.36 MHz accompanied with its 2nd harmonic. The electronic noise floor is also displayed in Fig. 2c (gray curve). Accordingly, the oscillation period corresponds to about 19 cavity roundtrips. We note that the oscillation period is not exactly an integer multiple of the roundtrip time, the latter phenomenon being called subharmonic entrainment and requiring a careful search of specific system parameters \cite{soto2005soliton,cole2019subharmonic,xian2020subharmonic}. 

\begin{figure*}
	\begin{center}
		\includegraphics[width=13.6cm]{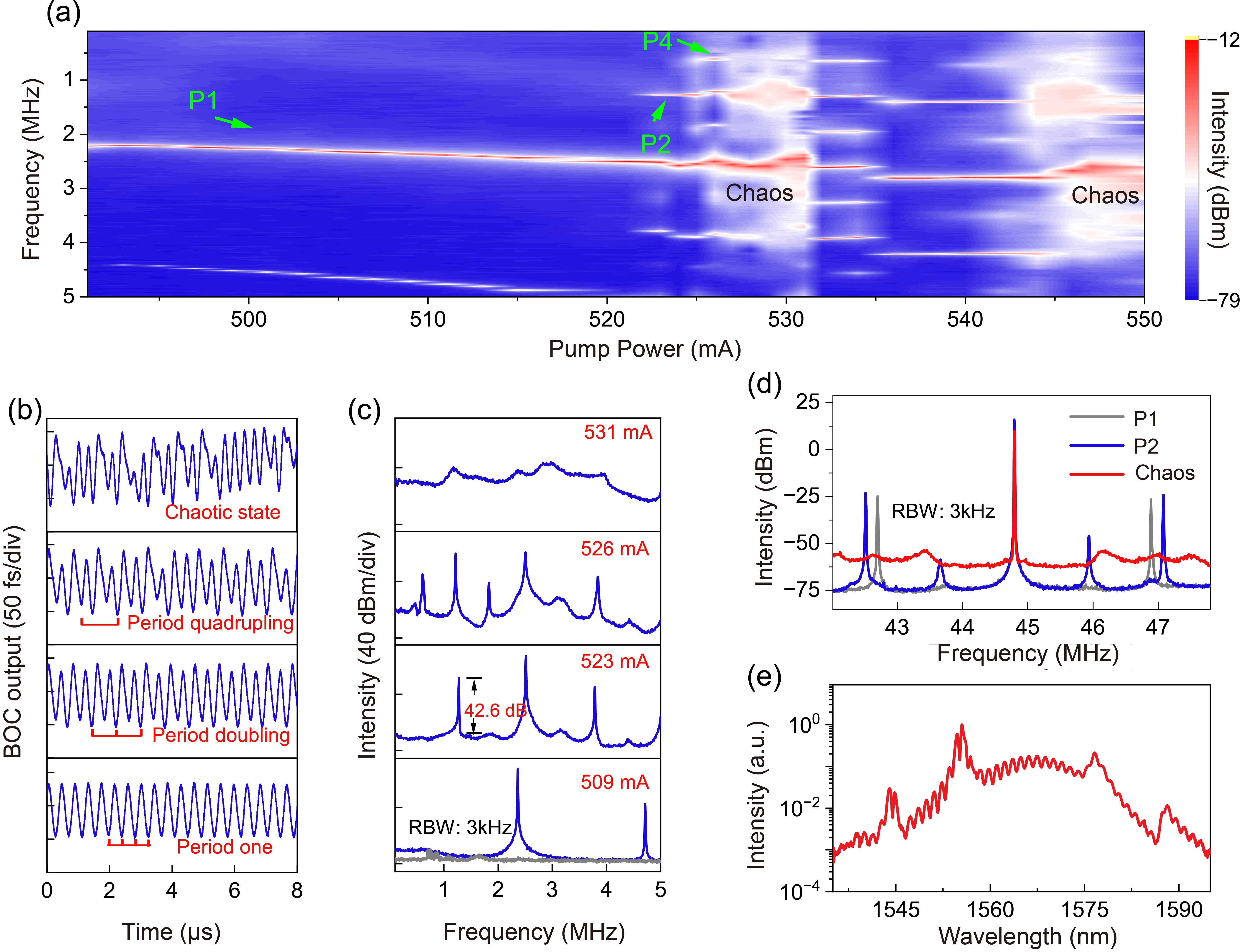}
		\caption{Generation of chaotic soliton molecules in an ultrafast fiber laser. (a) A full-record spectrogram shows the route of period doubling bifurcation to chaotic soliton molecules by increasing the pump strength. (b) Vibration waveform of the pulses separation for P1, P2, P4 and chaotic soliton molecules. (c) RF spectrum of the vibration waveforms displayed in (b). (d) RF spectrum of the laser output for P1, P2 and chaotic soliton molecules. (e) Average optical spectrum of a chaotic state recorded by an optical spectrum analyzer.}
	\end{center}
\end{figure*}

When the pump strength is increased, the P1 branch first maintains its stability with a slightly varying fundamental frequency, until a new set of spectral peaks appear at odd multiples of $f_{s}/2$ in the RF spectrogram (Fig. 2a), signaling a period-doubling bifurcation to a new family of period-doubled (P2) states. This bifurcation manifests in the vibration observed in the time domain (Fig. 2b) as a breaking of the symmetry of the oscillation pattern, which now repeats every two oscillations. We display the RF spectrum of the P2 state at $I$ $=$ 523 mA in Fig. 2c. The signal to noise ratio (SNR) of the $f_{s}/2$ frequency component is 42.6 dB, indicating a high stability. With a further increase of the pump driving current, the next bifurcation to the period-four (P4) family of steady states appears at $I$ $\approx$ 525 mA, see the vibration waveform in Fig. 2b and its RF spectrum in Fig. 2c. Finally, when the pump driving current increases above 526 mA, the sharp spectral peaks disappear and the RF spectrum becomes broadband, suggesting that the internal motion of soliton molecules has become chaotic. The two pulses, still bounded within the soliton molecule, experience irregular oscillations and their separation trajectory is aperiodic, in contrast with the periodic trajectories of the P1, P2, and P4 families. Here, a clear P1-P2-P4-…-chaos sequence is demonstrated, following the classical route of period-doubling bifurcation cascade to chaos. However, at a pump driving current exceeding 532 mA, the soliton molecule becomes once again periodic, first with limit cycle attractors belonging to a second P4 family, and then undergoes a reverse period-doubling bifurcation to reach a second P2 family. Finally, another transition to chaos starts for $I$ \textgreater  538 mA. In Fig. 2(d), we plot the RF spectrum of the direct photodetection of the laser output: it yields the cavity repetition frequency $f_{c}$, where both P1 oscillations (gray curve), P2 oscillations (blue curve) and chaos (red curve) are considered for comparison. For the P1 state, the fundamental cavity repetition frequency $f_{c}$ is accompanied by sharp modulation sidebands, whose separation from $f_{c}$ indeed matches with the BOC-measured oscillation frequency of the soliton molecule. 

Nevertheless, we emphasize that the relative amplitude of these sidebands remains small, typically 35 dB or more below the main $f_{c}$ peak: this indicates a minor effect of the internal motion of the soliton molecule on the total energy of the soliton molecule. This observation is important to rule out any soliton shape instability as a significant contributor to the chaos observed in our experiments \cite{soto2007soliton}. For the P2 state, a pair of sub-sidebands appears, with a frequency difference from $f_{c}$ being half of the initial P1 oscillating frequency. When reaching the chaotic states, the sharp modulation peaks disappear while the floor around $f_{c}$ increases by more than 10 dB compared with the P1 state, further demonstrating the aperiodicity of the chaotic state. 

\begin{figure*}
	\begin{center}
		\includegraphics[width=11cm]{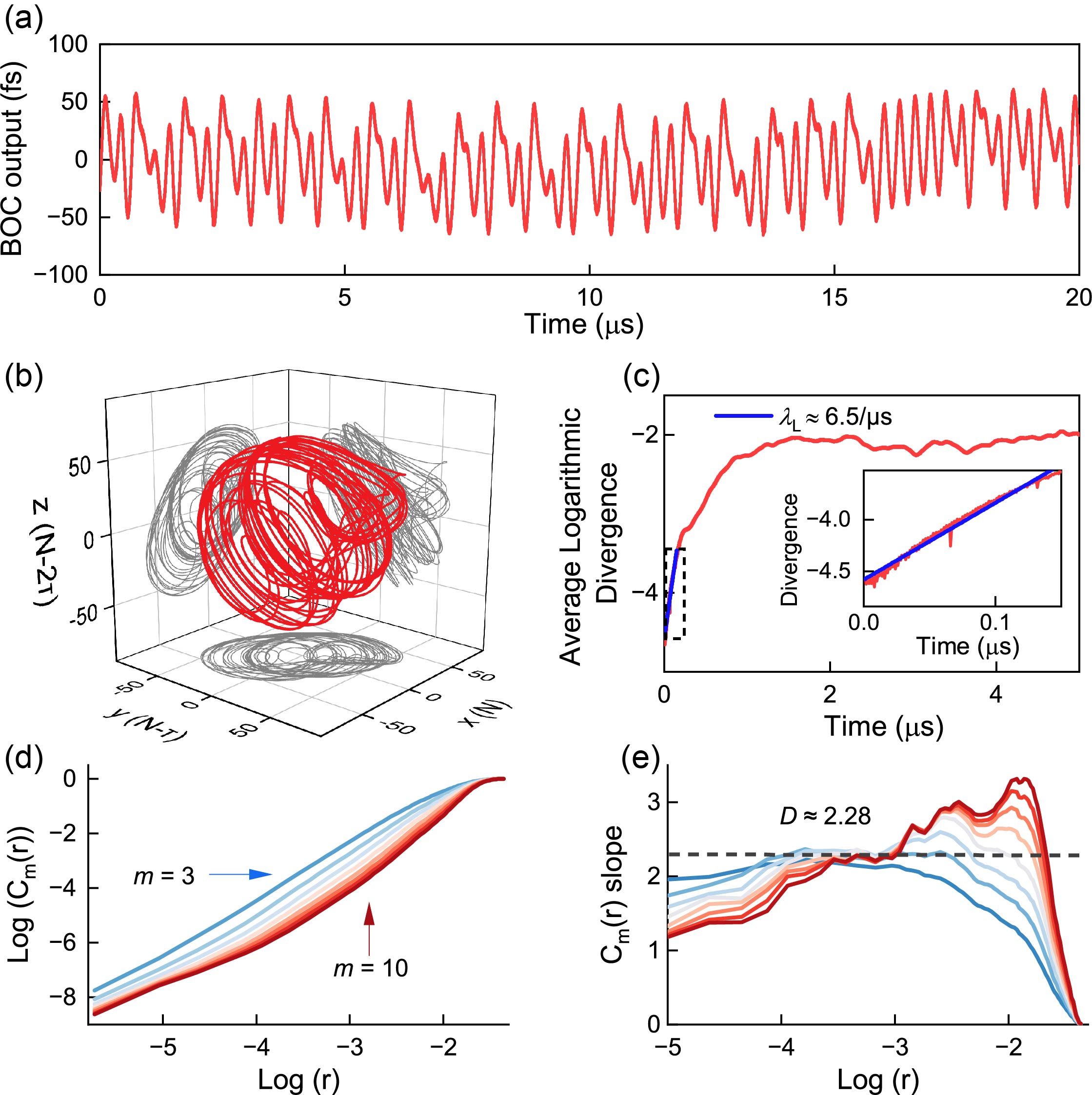}
		\caption{Statistical analysis of a chaotic internal soliton molecule motion. (a) BOC output signal. (b) A three-dimensional phase space orbit reconstructed from the BOC output signal and the delayed BOC output signals with delay times  $\tau$ and $2 \tau$. The gray curves are projections of the phase portrait on three phase planes. (c) Lyapunov exponent analysis of the BOC output signal. Inset: enlarged picture shows the exponent fitting result. (d) Logarithmic plot of the correlation function $C_{m}$($r$) versus the radius $r$ for increasing embedded dimension $m$. (e) Slope of the correlation function versus the radius $r$. A clear plateau is observed, with the black dashed curve giving an estimation of the correlation dimension.}
	\end{center}
\end{figure*}

As a highlighted feature of the internal soliton molecular motions reported here, from periodic to chaotic, we emphasize again on the fact that the optical soliton molecule always maintains its integrity, as a strongly bound soliton-pair system. This is linked to the fact that the relative timing variations, typically in the range of $\sim$ 100 fs, remain small compared to the average soliton separation, which is here about 6.4 ps and matches with the spectral interfringe of periodicity 1.28 nm shown in the average optical spectrum in Fig. 2(e). Finally, to support the view that these results reveal widely represented soliton molecule dynamics and are not a coincidence resulting from overly specific laser parameters, we have reproduced the bifurcation route to chaos of the internal motion of soliton molecules from a fiber laser cavity featuring a shorter SMF fiber length, see supplementary information Section B. We have also implemented a numerical simulation to qualitatively account for the observed pulse dynamics, based on the generalized nonlinear Schrödinger propagation equation, see supplementary information section C. The numerical results support our experimental findings of a clear route of period doubling bifurcation to chaos for the internal motion within soliton-pair molecules. In	 spite	 of	 these major bifurcations experienced by	the soliton	 molecule, we verify that the associated	intracavity	energy	fluctuations	remain	small ($\sim$1\% or below) see Section D, confirming	 the	 leading role of the internal	dynamics on	the	observed	phenomena.	.

\begin{figure*}
	\begin{center}
		\includegraphics[width=11.5cm]{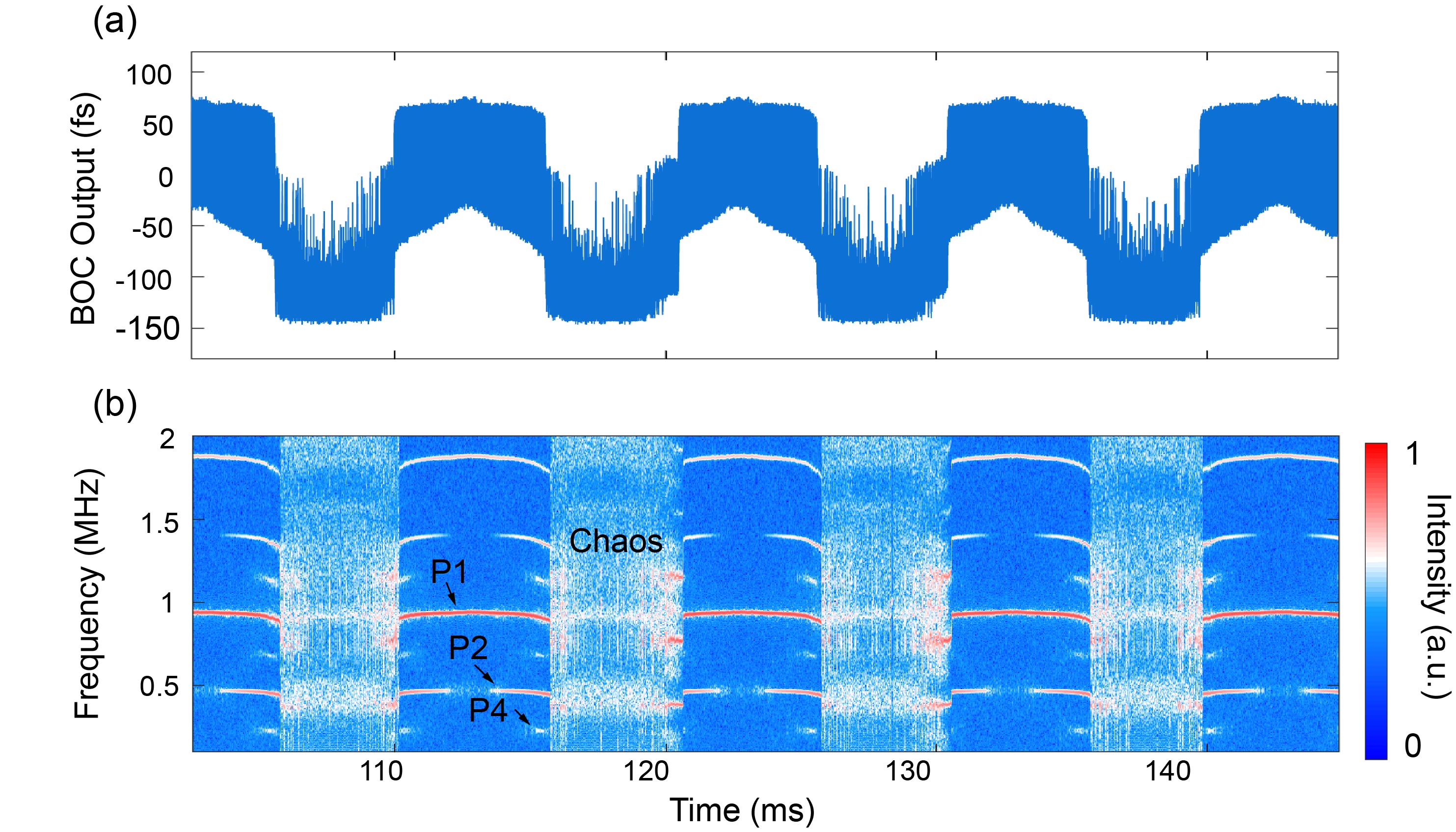}
		\caption{(a) Real-time BOC output signals of the variations of pulse separation within a soliton molecule during the pump current modulation at 100 Hz. (b) The two-dimensional spectra extracted from windowed Fourier transforms of the BOC output signals in (a).}
	\end{center}
\end{figure*}

\subsection{Dynamical characterization of chaotic soliton molecules}

Having established a deterministic route to chaotic soliton molecules, we next analyze the statistics of the chaotic trajectories recorded as in Fig. 2b, an example of which is shown in Fig. 3a. Figure 3b shows the chaotic orbit in a reconstructed three-dimensional phase space, which is obtained from the BOC output signal of Fig. 3a and the delayed BOC signals with delay times $\tau$ and 2$\tau$. The delay time $\tau$ is chosen by the mutual information method, which is frequently used in the phase space reconstruction of one-dimensional time series \cite{fraser1986independent}. The geometric structures can be seen from the projections of the phase trajectory on the x-y, x-z and y-z phase planes (gray curves). To quantitatively demonstrate that the reported interacting dynamics within the soliton molecule is chaotic, we extract the largest Lyapunov exponent $\lambda_{L}$ from the BOC signals. Lyapunov exponents quantify the divergence rate of nearby attractor trajectories in phase space and are widely used as a criterion for identifying chaos \cite{rosenstein1993practical}. A necessary condition for chaos is the existence of at least one positive Lyapunov exponent $\lambda_{L}$. As shown in Fig. 3c, the indicator of chaos is reflected by the average initial logarithmic divergence curve over all pairs of neighbors, where an exponential divergence is reasonably linear with the Lyapunov exponent $\lambda_{L}$ $\approx$ 6.5/$\mu$s. We note that the divergence trajectory features saturation since the chaotic system is bounded in the phase space so that the average divergence cannot exceed the boundary of the strange attractor. 

We also analyze the correlation dimension $D_{2}$ by using the Grassberger-Procaccia (GP) algorithm \cite{grassberger2004measuring}. The correlation dimension is an efficient diagnostic to distinguish between chaos and random motion. For a random motion, the measured $D_{2}$ grows linearly with the increasing embedding dimension $m$, whereas in the situation of chaos, $D_{2}$ saturates to an asymptotic value. Figure 3d displays the logarithmic plot of the correlation function $C_{m}$($r$) as a function of the radius $r$ for increasing $m$ and the corresponding correlation integral slope versus the radius $r$ is shown in Fig. 3e. The curve manifests a clear saturation of the correlation integral slope, towards an estimated correlation dimension $D_{2}$ $\approx$ 2.28. A finite value for the correlation dimension supports the hypothesis of chaotic internal soliton-pair dynamics and gives an estimate of the degree of complexity of the motion of the interacting solitons. A detailed description of the analyzing method involving the delay time $\tau$, the embedding dimension $m$, the Lyapunov exponent  $\lambda_{L}$, and the correlation dimension $D_{2}$ is presented in the Suppl. Section E. 

We verified the reproducibility of chaotic soliton molecules generation, by using a pump current modulated at 100 Hz with a wide modulation amplitude such that	the soliton molecules experience	a complete	circular	sequence of	bifurcations (P1$-$chao$s-$P1), as	observed in	Fig.	2a for	a driving	current	ranging from 500 to	530 mA. The resulting real-time BOC output signals of the pulse separation are shown in Fig. 4a, with Fourier spectral analysis in Fig. 4b: the recording demonstrates a perfect reproducibility of the bifurcation sequences leading to or departing from chaos. During the first half modulation period, switching occur from stable P1 to P2 state. A transient P4 state is then observed before the onset of chaos. Subsequently, the system returns to the P1 state through the reverse sequence of bifurcations, initiated by the	reduction of the intracavity energy. This dynamical switching between regular and chaotic oscillation is also illustrated in the Supplementary video 1 \& 2, where we videotaped this reversible process, showing the real-time evolution of waveforms and RF spectrograms after setting a longer modulation period for a clear view in real time.  Whereas the bifurcation sequence is repeatedly observed, the details of intra-molecular chaotic dynamics can vary substantially over consecutive experimental runs, with another example mirroring the analysis of Fig. 3 shown in the Supplemental figure S5. 

\begin{figure*}
	\begin{center}
		\includegraphics[width=12.8cm]{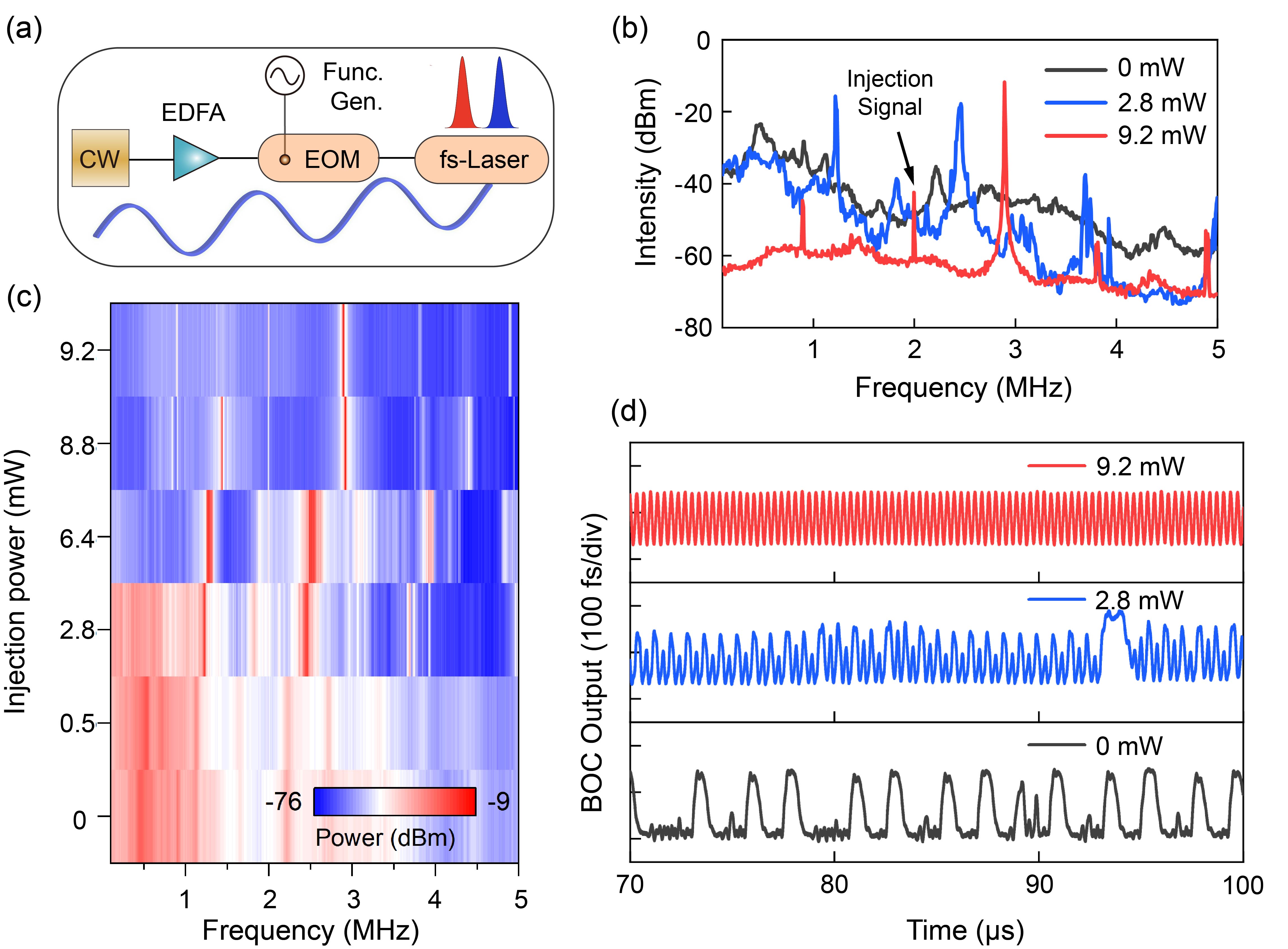}
		\caption{Statistical analysis of a chaotic internal soliton molecule motion. (a) BOC output signal. (b) A three-dimensional phase space orbit reconstructed from the BOC output signal and the delayed BOC output signals with delay times  $\tau$ and $2 \tau$. The gray curves are projections of the phase portrait on three phase planes. (c) Lyapunov exponent analysis of the BOC output signal. Inset: enlarged picture shows the exponent fitting result. (d) Logarithmic plot of the correlation function $C_{m}$($r$) versus the radius $r$ for increasing embedded dimension $m$. (e) Slope of the correlation function versus the radius $r$. A clear plateau is observed, with the black dashed curve giving an estimation of the correlation dimension.}
	\end{center}
\end{figure*}

\subsection{All optical control of chaotic soliton molecules}

Chaotic dynamics is prone to exhibit considerable reaction to additional perturbations. In chaos theory, it was shown that chaos can be suppressed to a limited cycle attractor by non-feedback methods such as the application of external periodic forces \cite{boccaletti2000control}. The elimination of chaos is most effective when the drive frequency (frequency of the external force) is close to the eigenfrequency of self-excited oscillations that exist in the vicinity (in the parameter space) of the chaotic motion. In our experiment, the free-oscillation frequency of the vibrating soliton molecule varies within the range of 1-5 MHz, so we choose a drive frequency $f_{d}$ = 2 MHz. We here demonstrate that the chaotic dynamics involving bound solitons in ultrafast fiber lasers can be suppressed by external cavity injection, without affecting the original laser parameters and the structure of the individual solitons. First, we experimentally verify that the soliton molecule remains a robust entity with respect to the injection signal, including within the chaotic region. Therefore, the soliton molecule presents an opportunity of external control. As shown in Fig. 5a, we implement the external control of chaos via the injection of a modulated continuous wave (CW), which is an efficient method for pulse dynamics modulation \cite{bao2015observation}. In the optical spectral domain, the CW is selected at 1530 nm, far enough from the soliton molecule spectrum to avoid a deleterious influence on the mode locking regime within a suitable range of injection power. The CW is amplified by an Er-doped fiber amplifier (EDFA), whose output power is adjustable. An electro optic modulator with a bandwidth exceeding 1 GHz driven by a function generator imprints a sinusoidal intensity modulation on the injection signal. The latter is injected into the cavity via a 10$\%$ coupler in the counterpropagating lasing direction, thus directly influencing the gain saturation level of the fiber laser with a bandwidth up to several MHz.

We experimentally prepare a chaotic soliton molecule whose dynamics of inter-pulse separation is manifestly broadband, as attested by the RF spectrum of the BOC output signal (black curve in Fig. 5b) at 550 mA pump strength. Then, we control this state by increasing the injected CW laser intensity, measured as the driving power of the external EDFA. At a driving power of 2.8 mW, the chaotic motion becomes intermittent, interspersed by P2-like oscillations (blue curve in Fig. 5d) with corresponding spectral peaks at half-integer multiples of $f_{s}$ located at about 2.5 MHz on top of a broad continuum (blue curve in Fig. 5b). Robust P1 oscillations are obtained at the current drive of 9.2 mW, as attested by both the RF spectrum and the regular vibration waveform (red curves in Figs. 5b \& d). The complete and deterministic route of chaos control through period-doubling reversals via external cavity injection is shown in Fig. 5c. Note that excessive injection power (\textgreater 9.8 mW) will result in the disruption of the soliton molecule. The frequency components of the injection signal and beat notes are also evidenced by the RF spectrum in Fig. 5b, confirming that the external injection is efficiently coupled to the chaotic motion. Overall, the route of external injection for chaos control evolves in a highly reproducible and reversible manner, so that the internal motion of soliton molecules reverts to chaos if we weaken the injection strength. For a clearer visual presentation of our results, we also videotaped this fully efficient switching of chaos control, see Section S3, Supplementary materials and video3.

\section{Discussion}

Whereas oscillating optical soliton molecules have been the subject of several seminal investigations, this article presents the first experimental demonstration of the route for excitation and all optical control of a chaotic soliton molecule in optical resonators and lasers. We emphasize the fact that the chaos investigated here refers to the internal dynamics of the soliton molecule and is assessed to be low-dimensional, leaving the soliton molecule energy practically unchanged. This can be understood physically by the chosen situation where the variations of the inter-pulse separation are of the order of one percent of that separation, therefore perturbing only slightly the overall shape of the soliton molecule. Therefore, to observe such internal chaos requires a characterization method endowed with an exquisite sub-femtosecond sensitivity and true real-time monitoring. The BOC method fully qualifies, doing so without relying on any heavy post-processing of data. This allowed us to record experimentally the vibration waveform with the necessary high precision to observe and analyze the experimental internal chaos. We therefore demonstrated that the chaotic dynamics of bound pulses within a soliton molecule can be accessed and manipulated in an unambiguous way. The BOC method appears particularly suited in the investigation of the complex nonlinear dynamics of soliton molecules in optical oscillators, as much as it is widely used for timing jitter characterization in ultrafast lasers and ring resonators \cite{benedick2012optical,bao2021quantum}. We anticipate that it will be a useful asset in precise investigation of nonlinear dynamics effects such as stochastic resonance \cite{monifi2016optomechanically} and subharmonic entrainment \cite{xian2020subharmonic} of solitons molecules.

Following the present investigation, the intramolecular dynamics of optical soliton molecules is gaining diversity as it further strengthens the analogy with matter-like molecules. Indeed, our study constitutes a strong confirmation of the hypothesis that a wide range of soliton molecule oscillations can be viewed as low-dimensional dynamical systems that feature limit cycle attractors as well as strange attractors. This hypothesis remains reasonably valid as long as the amplitude of the intramolecular motions remain sufficiently small compared to the average separation between solitons. 

The fast error-free switching between ordered and chaotic soliton molecules enabled by pump current sweeping and external injection highlights the potential prospects of all-optical logic gates and chaotic communication using soliton molecules \cite{pang2016all}. Beyond the fundamental case of soliton-pair molecules, there is a large variety of molecular complexes, macro-molecules and soliton crystals that could be similarly investigated. It is reasonable to assume that the chaotic dynamics observed in our experiment could be extended to three-soliton interactions in lasers \cite{akhmediev2005dissipative,xin2021evidence}, and even more complicated multi-body dynamics such as interactions within soliton molecular complexes \cite{wang2019optical} and supramolecular structures in parallel optical-soliton reactors \cite{he2021synthesis}. More generally, the findings of chaotic interactions among dissipative solitons will be of significant interest in the context of analogous nonlinear systems, such as Bose-Einstein condensates and hydrodynamics.

\begin{acknowledgments}
	The authors thank Dr. Dongyu Yan and Na Xiao for helpful discussions. This work is supported by the National Natural Science Foundation of China (Grant 61975144, 61827821); Ph.G. acknowledges support from the EIPHI Graduate School (ANR-17-EURE-0002) and from PIA3ISITE-BFC (ANR-15-IDEX-0003); O. Gat acknowledges support from the Israel Science Foundation (ISF) (Grant No. 2403/20).
\end{acknowledgments}

\appendix

\section{Laser setup}
The chaotic soliton molecules under investigation are generated from an Er-doped fiber laser, mode locked by the nonlinear polarization evolution (NPE) technique. NPE provides a virtual quasi-instantaneous saturable absorber effect, whose transfer function can be widely adjusted by altering the orientations of the intracavity wave plates, leading to multifarious mode-locked states. The cavity incorporates two types of fibers: a 0.5-m long EDF (Liekki Er 110-4/125) serves as the gain medium and the others are standard single-mode fiber (SMF 28). By changing the length of the SMF, we obtained two mode-locking states with different fundamental repetition frequency and net chromatic dispersion. The total length of the laser cavity of 4.6 (4.1) m yields a fundamental repetition frequency of 44.8 (50.6) MHz, corresponding to a roundtrip time of 22.3 (19.76) ns. The laser has a net anomalous dispersion $\sim$ -0.07(-0.06) ps$^{2}$ at 1.55-$\mu$m wavelength. In both states, the mode locking threshold is at a pump driving current of about 380 mA (optical pump power $\sim$ 230 mW) and the threshold of multi-pulsing (MPI) is at about 455 mA (optical power $\sim$ 270 mW). Soliton molecular chaotic dynamics are observed in both mode-locking states, proving that the observation is not a coincidence resulting from specific laser parameters.

\section{Balanced optical cross-correlator}
A single-crystal balanced optical cross-correlation (BOC) scheme has been implemented to detect the internal soliton molecular motion. The BOC setup consists of a 4-mm type-II phase-matched periodically poled KTiOPO$_{4}$ (PPKTP) crystal (facet 1 coating: HT @ 1550 nm \& 775 nm, facet 2 coating: HT @ 775 nm \& HR @ 1550 nm), a dichroic mirror (HT @1550 nm \& HR @ 775 nm), a focusing lens ($f$ $=$ 30 mm) and a balanced photodetector (Thorlabs PDB420A). The output of the BOC is detected by an oscilloscope (Agilent infiniium) for monitoring of intramolecular pulse separation in real time and a RF spectrum analyzer (Rigol DSA815) for spectrum analysis of chaotic dynamics. 

\section{Pump current modulation}
The pump laser diode (LD) is driven by a commercial laser diode controller (Thorlabs CLD1015), which has an external RF input port for pump current modulation. A 100 Hz sinusoidal modulation generated by a function generator (Rigol DG5102) is used to modulate the laser. The modulation amplitude can be adjusted by varying RF voltage from the function generator output. The modulation bandwidth can be as high as 10 MHz, while, in Er-fiber lasers, it is limited to few kilohertz due to the low pass filtering effect caused by the damped gain relaxation oscillations. Ref. 51 used this scheme for intramolecular pulse separation manipulation. Here, this technique allow us to reproduce the bifurcation sequences leading to or departing from chaos. 

\section{All-optical control of chaotic soliton molecules}
The external optical injection is based on an all-fiber configuration. The 1530 nm CW laser is generated by a commercial low noise tunable laser (Santec TSL-550). The output is power amplified by a home-made all polarization maintaining EDFA. A fiber-coupled high speed electro optic modulator (Conquer KG-AM-15-2.5G) is used to sinusoidally modulate the intensity of the CW and the MHz modulation signal is from a function generator (Rigol DG5102).

\bibliography{literature}

\providecommand{\href}[2]{#2} \providecommand{\beforedoihref}{}
  \providecommand{\afterdoihref}{}\begingroup\raggedright\begin{thebibliography}{10}

\bibitem{poincare1890probleme}
H.~Poincar{\'e}, {\it Sur le probl{\`e}me des trois corps et les {\'e}quations
  de la dynamique},  Acta mathematica {\bf 13} (1890), no.~1 A3--A270.

\bibitem{lorenz1963deterministic}
E.~N. Lorenz, {\it Deterministic nonperiodic flow},  Journal of atmospheric
  sciences {\bf 20} (1963), no.~2 130--141.

\bibitem{strogatz2018nonlinear}
S.~H. Strogatz, {\em Nonlinear dynamics and chaos: with applications to
  physics, biology, chemistry, and engineering}.
\newblock CRC press, 2018.

\bibitem{argyris2005chaos}
A.~Argyris, D.~Syvridis, L.~Larger, V.~Annovazzi-Lodi, P.~Colet, I.~Fischer,
  J.~Garcia-Ojalvo, C.~R. Mirasso, L.~Pesquera and K.~A. Shore, {\it
  Chaos-based communications at high bit rates using commercial fibre-optic
  links},  Nature {\bf 438} (2005), no.~7066 343--346.

\bibitem{uchida2008fast}
A.~Uchida, K.~Amano, M.~Inoue, K.~Hirano, S.~Naito, H.~Someya, I.~Oowada,
  T.~Kurashige, M.~Shiki, S.~Yoshimori {et.~al.}, {\it Fast physical random bit
  generation with chaotic semiconductor lasers},  Nat. Photonics {\bf 2}
  (2008), no.~12 728--732.

\bibitem{spitz2021private}
O.~Spitz, A.~Herdt, J.~Wu, G.~Maisons, M.~Carras, C.-W. Wong,
  W.~Els{\"a}{\ss}er and F.~Grillot, {\it Private communication with quantum
  cascade laser photonic chaos},  Nat. Commun. {\bf 12} (2021), no.~1 1--8.

\bibitem{sciamanna2015physics}
M.~Sciamanna and K.~A. Shore, {\it Physics and applications of laser diode
  chaos},  Nat. Photonics {\bf 9} (2015), no.~3 151--162.

\bibitem{jiang2017chaos}
X.~Jiang, L.~Shao, S.-X. Zhang, X.~Yi, J.~Wiersig, L.~Wang, Q.~Gong,
  M.~Lon{\v{c}}ar, L.~Yang and Y.-F. Xiao, {\it Chaos-assisted broadband
  momentum transformation in optical microresonators},  Science {\bf 358}
  (2017), no.~6361 344--347.

\bibitem{jumpertz2016chaotic}
L.~Jumpertz, K.~Schires, M.~Carras, M.~Sciamanna and F.~Grillot, {\it Chaotic
  light at mid-infrared wavelength},  Light Sci. Appl. {\bf 5} (2016), no.~6
  e16088--e16088.

\bibitem{fan2021real}
L.~Fan, X.~Yan, H.~Wang and L.~V. Wang, {\it Real-time observation and control
  of optical chaos},  Science advances {\bf 7} (2021), no.~3 eabc8448.

\bibitem{deng2022mid}
Y.~Deng, Z.-F. Fan, B.-B. Zhao, X.-G. Wang, S.~Zhao, J.~Wu, F.~Grillot and
  C.~Wang, {\it Mid-infrared hyperchaos of interband cascade lasers},  Light
  Sci. Appl. {\bf 11} (2022), no.~1 1--10.

\bibitem{steinmeyer2021chaotic}
G.~Steinmeyer and W.~Chen, {\it Chaotic synchronization in optical frequency
  combs},  arXiv preprint arXiv:2110.10045 (2021).

\bibitem{grelu2012dissipative}
P.~Grelu and N.~Akhmediev, {\it Dissipative solitons for mode-locked lasers},
  Nat. Photonics {\bf 6} (2012), no.~2 84--92.

\bibitem{sucha1995period}
G.~Sucha, S.~Bolton, S.~Weiss and D.~Chemla, {\it Period doubling and
  quasi-periodicity in additive-pulse mode-locked lasers},  Opt. Lett. {\bf 20}
  (1995), no.~17 1794--1796.

\bibitem{xing1999regular}
Q.~Xing, L.~Chai, W.~Zhang and C.-Y. Wang, {\it Regular, period-doubling,
  quasi-periodic, and chaotic behavior in a self-mode-locked ti: sapphire
  laser},  Opt. Comm. {\bf 162} (1999), no.~1-3 71--74.

\bibitem{soto2004bifurcations}
J.~M. Soto-Crespo, M.~Grapinet, P.~Grelu and N.~Akhmediev, {\it Bifurcations
  and multiple-period soliton pulsations in a passively mode-locked fiber
  laser},  Phys. Rev. E {\bf 70} (2004), no.~6 066612.

\bibitem{zhao2006chaotic}
L.~Zhao, D.~Tang and A.~Liu, {\it {Chaotic dynamics of a passively mode-locked
  soliton fiber ring laser}},  Chaos: An Interdisciplinary Journal of Nonlinear
  Science {\bf 16} (2006), no.~1 013128.

\bibitem{cundiff2002experimental}
S.~T. Cundiff, J.~M. Soto-Crespo and N.~Akhmediev, {\it Experimental evidence
  for soliton explosions},  Phys. Rev. Lett. {\bf 88} (2002), no.~7 073903.

\bibitem{runge2015observation}
A.~F. Runge, N.~G. Broderick and M.~Erkintalo, {\it Observation of soliton
  explosions in a passively mode-locked fiber laser},  Optica {\bf 2} (2015),
  no.~1 36--39.

\bibitem{horowitz1997noiselike}
M.~Horowitz, Y.~Barad and Y.~Silberberg, {\it Noiselike pulses with a broadband
  spectrum generated from an erbium-doped fiber laser},  Opt. Lett. {\bf 22}
  (1997), no.~11 799--801.

\bibitem{runge2013coherence}
A.~F. Runge, C.~Aguergaray, N.~G. Broderick and M.~Erkintalo, {\it Coherence
  and shot-to-shot spectral fluctuations in noise-like ultrafast fiber lasers},
   Opt. Lett. {\bf 38} (2013), no.~21 4327--4330.

\bibitem{lecaplain2014rogue}
C.~Lecaplain and P.~Grelu, {\it Rogue waves among noiselike-pulse laser
  emission: an experimental investigation},  Phys. Rev. A {\bf 90} (2014),
  no.~1 013805.

\bibitem{krupa2017vector}
K.~Krupa, K.~Nithyanandan and P.~Grelu, {\it Vector dynamics of incoherent
  dissipative optical solitons},  Optica {\bf 4} (2017), no.~10 1239--1244.

\bibitem{soto2005soliton}
J.~M. Soto-Crespo and N.~Akhmediev, {\it Soliton as strange attractor:
  nonlinear synchronization and chaos},  Phys. Rev. Lett. {\bf 95} (2005),
  no.~2 024101.

\bibitem{wang2020buildup}
Z.~Wang, K.~Nithyanandan, A.~Coillet, P.~Tchofo-Dinda and P.~Grelu, {\it
  Buildup of incoherent dissipative solitons in ultrafast fiber lasers},  Phys.
  Rev. Research {\bf 2} (2020), no.~1 013101.

\bibitem{lucas2017breathing}
E.~Lucas, M.~Karpov, H.~Guo, M.~Gorodetsky and T.~J. Kippenberg, {\it Breathing
  dissipative solitons in optical microresonators},  Nat. Commun. {\bf 8}
  (2017), no.~1 1--11.

\bibitem{melo2018deterministic}
L.~B. M{\'e}lo, G.~F. Palacios, P.~V. Carelli, L.~H. Acioli, J.~R.~R. Leite and
  M.~H. de~Miranda, {\it Deterministic chaos in an ytterbium-doped mode-locked
  fiber laser},  Opt. Express {\bf 26} (2018), no.~10 13686--13692.

\bibitem{wang2022real}
W.~Wang, T.~Xian, M.~Zhang and L.~Zhan, {\it Real-time observation of phase
  transition of bifurcation evolution in mode-locked lasers},  Opt. Lett. {\bf
  47} (2022), no.~5 1234--1237.

\bibitem{shlizerman2011characterizing}
E.~Shlizerman, E.~Ding, M.~O. Williams and J.~N. Kutz, {\it Characterizing and
  suppressing multi-pulsing instabilities in mode-locked lasers},  in {\em
  Physics and Simulation of Optoelectronic Devices XIX}, vol.~7933,
  pp.~327--341, SPIE, 2011.

\bibitem{lecaplain2012dissipative}
C.~Lecaplain, P.~Grelu, J.~Soto-Crespo and N.~Akhmediev, {\it Dissipative rogue
  waves generated by chaotic pulse bunching in a mode-locked laser},  Phys.
  Rev. Lett. {\bf 108} (2012), no.~23 233901.

\bibitem{chouli2010soliton}
S.~Chouli and P.~Grelu, {\it Soliton rains in a fiber laser: An experimental
  study},  Phys. Rev. A {\bf 81} (2010), no.~6 063829.

\bibitem{zaviyalov2012rogue}
A.~Zaviyalov, O.~Egorov, R.~Iliew and F.~Lederer, {\it Rogue waves in
  mode-locked fiber lasers},  Phys. Rev. A {\bf 85} (2012), no.~1 013828.

\bibitem{liu2016successive}
M.~Liu, A.-P. Luo, Y.-R. Yan, S.~Hu, Y.-C. Liu, H.~Cui, Z.-C. Luo and W.-C. Xu,
  {\it Successive soliton explosions in an ultrafast fiber laser},  Opt. Lett.
  {\bf 41} (2016), no.~6 1181--1184.

\bibitem{meng2021intracavity}
F.~Meng, C.~Lapre, C.~Billet, T.~Sylvestre, J.-M. Merolla, C.~Finot, S.~K.
  Turitsyn, G.~Genty and J.~M. Dudley, {\it Intracavity incoherent
  supercontinuum dynamics and rogue waves in a broadband dissipative soliton
  laser},  Nat. Commun. {\bf 12} (2021), no.~1 1--12.

\bibitem{soto2003quantized}
J.~M. Soto-Crespo, N.~Akhmediev, P.~Grelu and F.~Belhache, {\it Quantized
  separations of phase-locked soliton pairs in fiber lasers},  Opt. Lett. {\bf
  28} (2003), no.~19 1757--1759.

\bibitem{tang2005mechanism}
D.~Tang, L.-M. Zhao, B.~Zhao and A.~Liu, {\it Mechanism of multisoliton
  formation and soliton energy quantization in passively mode-locked fiber
  lasers},  Phys. Rev. A {\bf 72} (2005), no.~4 043816.

\bibitem{wang2017universal}
Y.~Wang, F.~Leo, J.~Fatome, M.~Erkintalo, S.~G. Murdoch and S.~Coen, {\it
  Universal mechanism for the binding of temporal cavity solitons},  Optica
  {\bf 4} (2017), no.~8 855--863.

\bibitem{song2020attosecond}
Y.~Song, F.~Zhou, H.~Tian and M.~Hu, {\it Attosecond timing jitter within a
  temporal soliton molecule},  Optica {\bf 7} (2020), no.~11 1531--1534.

\bibitem{grapinet2006vibrating}
M.~Grapinet and P.~Grelu, {\it Vibrating soliton pairs in a mode-locked laser
  cavity},  Opt. Lett. {\bf 31} (2006), no.~14 2115--2117.

\bibitem{soto2007soliton}
J.~M. Soto-Crespo, P.~Grelu, N.~Akhmediev and N.~Devine, {\it Soliton complexes
  in dissipative systems: Vibrating, shaking, and mixed soliton pairs},  Phys.
  Rev. E {\bf 75} (2007), no.~1 016613.

\bibitem{ortacc2010observation}
B.~Orta{\c{c}}, A.~Zaviyalov, C.~K. Nielsen, O.~Egorov, R.~Iliew, J.~Limpert,
  F.~Lederer and A.~T{\"u}nnermann, {\it Observation of soliton molecules with
  independently evolving phase in a mode-locked fiber laser},  Opt. Lett. {\bf
  35} (2010), no.~10 1578--1580.

\bibitem{krupa2017real}
K.~Krupa, K.~Nithyanandan, U.~Andral, P.~Tchofo-Dinda and P.~Grelu, {\it
  Real-time observation of internal motion within ultrafast dissipative optical
  soliton molecules},  Phys. Rev. Lett. {\bf 118} (2017), no.~24 243901.

\bibitem{herink2017real}
G.~Herink, F.~Kurtz, B.~Jalali, D.~R. Solli and C.~Ropers, {\it Real-time
  spectral interferometry probes the internal dynamics of femtosecond soliton
  molecules},  Science {\bf 356} (2017), no.~6333 50--54.

\bibitem{liu2018real}
X.~Liu, X.~Yao and Y.~Cui, {\it Real-time observation of the buildup of soliton
  molecules},  Phys. Rev. Lett. {\bf 121} (2018), no.~2 023905.

\bibitem{wang2019optical}
Z.~Wang, K.~Nithyanandan, A.~Coillet, P.~Tchofo-Dinda and P.~Grelu, {\it
  Optical soliton molecular complexes in a passively mode-locked fibre laser},
  Nat. Commun. {\bf 10} (2019), no.~1 1--11.

\bibitem{melchert2019soliton}
O.~Melchert, S.~Willms, S.~Bose, A.~Yulin, B.~Roth, F.~Mitschke, U.~Morgner,
  I.~Babushkin and A.~Demircan, {\it Soliton molecules with two frequencies},
  Phys. Rev. Lett. {\bf 123} (2019), no.~24 243905.

\bibitem{luo2020real}
Y.~Luo, R.~Xia, P.~P. Shum, W.~Ni, Y.~Liu, H.~Q. Lam, Q.~Sun, X.~Tang and
  L.~Zhao, {\it Real-time dynamics of soliton triplets in fiber lasers},
  Photonics Research {\bf 8} (2020), no.~6 884--891.

\bibitem{zhou2020buildup}
Y.~Zhou, Y.-X. Ren, J.~Shi, H.~Mao and K.~K. Wong, {\it Buildup and
  dissociation dynamics of dissipative optical soliton molecules},  Optica {\bf
  7} (2020), no.~8 965--972.

\bibitem{peng2021breather}
J.~Peng, Z.~Zhao, S.~Boscolo, C.~Finot, S.~Sugavanam, D.~V. Churkin and
  H.~Zeng, {\it Breather molecular complexes in a passively mode-locked fiber
  laser},  Laser \& Photonics Rev. {\bf 15} (2021), no.~7 2000132.

\bibitem{du2022internal}
Y.~Du, Z.~He, Q.~Gao, C.~Zeng, D.~Mao and J.~Zhao, {\it Internal dynamics in
  bound states of unequal solitons},  Opt. Lett. {\bf 47} (2022), no.~7
  1618--1621.

\bibitem{nimmesgern2021soliton}
L.~Nimmesgern, C.~Beckh, H.~Kempf, A.~Leitenstorfer and G.~Herink, {\it Soliton
  molecules in femtosecond fiber lasers: universal binding mechanism and direct
  electronic control},  Optica {\bf 8} (2021), no.~10 1334--1339.

\bibitem{hamdi2018real}
S.~Hamdi, A.~Coillet and P.~Grelu, {\it Real-time characterization of optical
  soliton molecule dynamics in an ultrafast thulium fiber laser},  Opt. Lett.
  {\bf 43} (2018), no.~20 4965--4968.

\bibitem{hamdi2022superlocalization}
S.~Hamdi, A.~Coillet, B.~Cluzel, P.~Grelu and P.~Colman, {\it Superlocalization
  reveals long-range synchronization of vibrating soliton molecules},  Phys.
  Rev. Lett. {\bf 128} (2022), no.~21 213902.

\bibitem{kurtz2020resonant}
F.~Kurtz, C.~Ropers and G.~Herink, {\it Resonant excitation and all-optical
  switching of femtosecond soliton molecules},  Nat. Photonics {\bf 14} (2020),
  no.~1 9--13.

\bibitem{zou2022synchronization}
D.~Zou, Y.~Song, O.~Gat, M.~Hu and P.~Grelu, {\it Synchronization of the
  internal dynamics of optical soliton molecules},  arXiv preprint
  arXiv:2208.10076 (2022).

\bibitem{turaev2007chaotic}
D.~Turaev, A.~G. Vladimirov and S.~Zelik, {\it Chaotic bound state of localized
  structures in the complex ginzburg-landau equation},  Phys. Rev. E {\bf 75}
  (2007), no.~4 045601.

\bibitem{wu2005nonlinearity}
G.~Wu, {\em Nonlinearity and chaos in molecular vibrations}.
\newblock Elsevier, 2005.

\bibitem{kim2007attosecond}
J.~Kim, J.~Chen, J.~Cox and F.~X. K{\"a}rtner, {\it Attosecond-resolution
  timing jitter characterization of free-running mode-locked lasers},  Opt.
  Lett. {\bf 32} (2007), no.~24 3519--3521.

\bibitem{jia2020photonic}
K.~Jia, X.~Wang, D.~Kwon, J.~Wang, E.~Tsao, H.~Liu, X.~Ni, J.~Guo, M.~Yang,
  X.~Jiang {et.~al.}, {\it Photonic flywheel in a monolithic fiber resonator},
  Phys. Rev. Lett. {\bf 125} (2020), no.~14 143902.

\bibitem{ott2002chaos}
E.~Ott, {\em Chaos in dynamical systems}.
\newblock Cambridge university press, 2002.

\bibitem{cole2019subharmonic}
D.~C. Cole and S.~B. Papp, {\it Subharmonic entrainment of kerr breather
  solitons},  Phys. Rev. Lett. {\bf 123} (2019), no.~17 173904.

\bibitem{xian2020subharmonic}
T.~Xian, L.~Zhan, W.~Wang and W.~Zhang, {\it Subharmonic entrainment breather
  solitons in ultrafast lasers},  Phys. Rev. Lett. {\bf 125} (2020), no.~16
  163901.

\bibitem{fraser1986independent}
A.~M. Fraser and H.~L. Swinney, {\it Independent coordinates for strange
  attractors from mutual information},  Phys. Rev. A {\bf 33} (1986), no.~2
  1134.

\bibitem{rosenstein1993practical}
M.~T. Rosenstein, J.~J. Collins and C.~J. De~Luca, {\it A practical method for
  calculating largest lyapunov exponents from small data sets},  Physica D:
  Nonlinear Phenomena {\bf 65} (1993), no.~1-2 117--134.

\bibitem{grassberger2004measuring}
P.~Grassberger and I.~Procaccia, {\it Measuring the strangeness of strange
  attractors},  in {\em The theory of chaotic attractors}, pp.~170--189.
\newblock Springer, 2004.

\bibitem{boccaletti2000control}
S.~Boccaletti, C.~Grebogi, Y.-C. Lai, H.~Mancini and D.~Maza, {\it The control
  of chaos: theory and applications},  Physics reports {\bf 329} (2000), no.~3
  103--197.

\bibitem{bao2015observation}
C.~Bao, W.~Chang, C.~Yang, N.~Akhmediev and S.~T. Cundiff, {\it Observation of
  coexisting dissipative solitons in a mode-locked fiber laser},  Phys. Rev.
  Lett. {\bf 115} (2015), no.~25 253903.

\bibitem{benedick2012optical}
A.~J. Benedick, J.~G. Fujimoto and F.~X. K{\"a}rtner, {\it Optical flywheels
  with attosecond jitter},  Nat. Photonics {\bf 6} (2012), no.~2 97--100.

\bibitem{bao2021quantum}
C.~Bao, M.-G. Suh, B.~Shen, K.~{\c{S}}afak, A.~Dai, H.~Wang, L.~Wu, Z.~Yuan,
  Q.-F. Yang, A.~B. Matsko {et.~al.}, {\it Quantum diffusion of microcavity
  solitons},  Nat. Phys. {\bf 17} (2021), no.~4 462--466.

\bibitem{monifi2016optomechanically}
F.~Monifi, J.~Zhang, {\c{S}}.~K. {\"O}zdemir, B.~Peng, Y.-x. Liu, F.~Bo,
  F.~Nori and L.~Yang, {\it Optomechanically induced stochastic resonance and
  chaos transfer between optical fields},  Nat. Photonics {\bf 10} (2016),
  no.~6 399--405.

\bibitem{pang2016all}
M.~Pang, W.~He, X.~Jiang and P.~S.~J. Russell, {\it All-optical bit storage in
  a fibre laser by optomechanically bound states of solitons},  Nat. Photonics
  {\bf 10} (2016), no.~7 454--458.

\bibitem{akhmediev2005dissipative}
N.~Akhmediev, J.~M. Soto-Crespo, M.~Grapinet and P.~Grelu, {\it Dissipative
  soliton interactions inside a fiber laser cavity},  Optical Fiber Technology
  {\bf 11} (2005), no.~3 209--228.

\bibitem{xin2021evidence}
F.~Xin, F.~Di~Mei, L.~Falsi, D.~Pierangeli, C.~Conti, A.~J. Agranat and
  E.~DelRe, {\it Evidence of chaotic dynamics in three-soliton collisions},
  Phys. Rev. Lett. {\bf 127} (2021), no.~13 133901.

\bibitem{he2021synthesis}
W.~He, M.~Pang, D.-H. Yeh, J.~Huang, P.~Russell and J.~St, {\it Synthesis and
  dissociation of soliton molecules in parallel optical-soliton reactors},
  Light Sci. Appl. {\bf 10} (2021), no.~1 1--15.

\end{thebibliography}\endgroup
\bibliographystyle{bibstyl}

\end{document}